\documentclass[11pt,twoside]{article}

\usepackage{asp2006}
\usepackage{epsf}
\usepackage{amsmath,amsfonts,graphicx}
\usepackage{lscape}

\def\bE{{\bf E}}
\def\bB{{\bf B}}
\def\bv{{\bf v}}
\def\bx{{\bf x}}
\def\bJ{{\bf J}}

\begin{document}
\title{The algorithms of  the implicit method}   
\author{Giovanni Lapenta}   
\affil{Centrum voor
Plasma-Astrofysica, Departement Wiskunde,  Katholieke Universiteit
Leuven, Celestijnenlaan 200B - bus 2400, B-3001 Heverlee, Belgi\"e.}    

\begin{abstract} 
We discuss the fundamentals of the implicit moment method for Particle In Cell (PIC) simulation as presently implemented in the CELESTE3D code. We present the method in its fully electromagnetic and fully
kinetic version. The application of the method is to problems with multiple temporal and spatial scales, common in all space, astrophysical and laboratory plasmas.
\end{abstract}



\section{Introduction}

The implicit  particle in cell (PIC) method was developed as a
general plasma simulation tool based on a fully kinetic
approach~\citep{mason1981,denavit1981} that did not rely on any
physical approximation but only on the use of advanced numerical
methods to reduce the cost of large scale kinetic simulations. 

 In its simplest form the PIC method uses explicit time discretization methods~\citep{birdsall}.  The equations of motion need
the fields acting on the computational particles and
the field equations need the moments computed from the particles.
The great majority of PIC codes in use today address this problem
relying on a explicit method.

In the explicit method, the field equations are discretized with one
in  a number of different explicit methods available (see
~\citet{birdsall,hockney} for a review) and the equations of
motion are usually discretized with the leap-frog
algorithm~\citep{birdsall,hockney}. The key point is that in the
explicit method, the field equations need only the sources from the
previous time cycle and the equations of motions need only the
fields from the previous time cycle. Even though the equations
remain coupled, no iteration is needed and the cycle  becomes a simple marching order where each
block is applied after its predecessor and needs only information
already available. This choice makes the explicit PIC very simple.
As can be imagined this simplicity comes at a price. The explicit
PIC approach is subject to three very restrictive stability
constraints.

First, the explicit discretization of the field equations requires
that a Courant condition must be satisfied on the speed of light:
\begin{equation}
c \Delta t <\Delta x
\end{equation}

Second, the explicit discretization of the equations of motion
introduces a constraint related to the fastest electron response
time, the electron plasma frequency:
\begin{equation}
\omega_{pe} \Delta t <\Delta x
\end{equation}
Note that, instead, thanks to the Boris algorithm for the motion in
a magnetic field, the gyromotion introduces no stability
constraints~\citep{birdsall}.

Finally, the interpolation between grid and particles causes a loss
of information and an aliasing instability called finite grid
instability that results in an additional stability constraint:
\begin{equation}
\Delta x <\varsigma \lambda_{De}
\end{equation}
that requires the grid spacing to be of the order of the Debye
length or smaller, the proportionality constant $\varsigma$ being
of order one and dependent on the details of the scheme
used~\citep{birdsall,hockney}.

Thanks to ever faster computers, the explicit PIC method has been
able to achieve remarkable results. But there are problems where multiple scales still
prevent its use. When the interest is on large scales and slow processes other approaches are needed. The implicit PIC approach
provides a viable alternative in such cases.

\section{Implicit PIC}

The implicit PIC method has been developed along two different lines of investigation:
the direct implicit method~\citep{directimplicit} and the implicit
moment method~\citep{brackbill-forslund}.  Here we consider the
implicit moment method. In the last few years, we have published a
number of papers on the application of CELESTE3D and a full description of the methods used has been provided by~\cite{lapenta06}.
In the present work, we  summarize the latest status of the implicit moment method as implemented in CELESTE3D.

We consider first the time discretization. The solution is advanced
in time with discrete steps, $\Delta t$, from the
initial time, $t^0=0$ to the final time $t^N=T$. The generic
quantity $\Psi$ at time step $n$ ($t=t^n$),  is denoted with
$\Psi^n$.

The equations of motion are discretized as \citep{vu1992}:
\begin{equation}
\begin{array}{l}
\displaystyle \bx_p^{n+1}=\bx_p^{n}+\bv^{n+1/2}_p \Delta t \\
\\
\displaystyle \bv_p^{n+1}=\bv_p^{n}+\frac{q_s\Delta
t}{m_s}\left(\bE^{n+\theta}_p(\bx^{n+1/2}_p)+\bv^{n+1/2}_p\times
\bB^n_p(\bx^{n+1/2}_p)\right)
\end{array} \label{mover}
\end{equation}
where all quantities evaluated at intermediate levels are computed
as: $\Psi^{n+\theta}=\Psi^n(1-\theta)+\Psi^{n+1}\theta$. Note that
the velocity equation is more conveniently rewritten as:
\begin{equation}
\bv^{n+1/2}_p= \widehat{\bv}_p +\beta_s
\widehat{\bE}_p^{n+\theta}(\bx^{n+1/2}_p)
 \label{vel_eq}
\end{equation}
where $\beta_s=q_p \Delta t/m_p$ (independent of the particle weight
and unique to a given species). For convenience, we have introduced
hatted quantities obtained by explicit transformation of quantities
known from the previous computational cycle:
\begin{equation}
\begin{array}{c}
 \widehat{\bv}_p = \boldsymbol{\alpha}^n_s \cdot \bv^n_p \\ \\
\widehat{\bE}_s^{n+\theta} = \boldsymbol{\alpha}^n_s \cdot
\bE_s^{n+\theta}
\end{array}
\end{equation}
The transformation tensor operators $\boldsymbol{\alpha}_s^n$ are
defined as:
\begin{equation}
\boldsymbol{\alpha}_s^n =  \frac{1}{1+(\beta_s B^{n})^2}
\left(\boldsymbol{I}-\beta_s \boldsymbol{I} \times \bB^n +\beta_s^2
\bB^n \bB^n \right)
\end{equation}
and represent a scaling and rotation of the velocity vector.

The semi--discrete (continuous in space) temporal discretization to
Maxwell's equations is written as:
\begin{equation}
\label{differencedMAX}
\begin{array}{c}
   \nabla \times \mathbf{E}^{n+\theta} +
\displaystyle{\frac{1}{c}} \frac{\mathbf{B}^{n+1}-
\mathbf{B}^n}{\Delta t} =0 \\ \\
   \nabla \times \mathbf{B}^{n+\theta} - \displaystyle{\frac{1}{c}}
\frac{ \mathbf{E}^{n+1}-\mathbf{E}^n }{\Delta t}
    =\frac{4\pi}{c}\mathbf{J}^{n+\frac{1}{2}} \\ \\
   \nabla \cdot   \mathbf{E}^{n+\theta}
= 4 \pi \rho^{n+\theta} \\ \\
   \nabla \cdot \mathbf{B}^n=\nabla \cdot \mathbf{B}^{n+1}  = 0,
\end{array}
\end{equation}

The parameter $\theta \in [1/2,1]$ is chosen in order to adjust the
numerical dispersion relation for electromagnetic waves (for
$\theta<1/2$, the algorithm is shown to be unstable
\citep{brackbill-forslund}). We note that for $\theta=1/2$ the scheme
is second-order accurate in $\Delta t$; for $1/2 < \theta \le 1$ the
scheme is first-order accurate.

The sources in
Maxwell's equations necessitate information from the particles:
\begin{equation}
\begin{array}{c}
 \rho_{s} = \sum_{p =1}^{N_s} q_p W({\bf x}-{\bf
x}_p) \\ \\ {\bf J}_{s}({\bf r}) = \sum_{p =1}^{N_s} q_p {\bf v}_p
W({\bf x}-{\bf x}_p)
\end{array} \label{moments_int}
\end{equation}
where the species is labelled by $s$ and the sums are carried over all particles of a species $N_s$.
 The coupling with the particle equations of motion is
evident.

The fundamental problem to address in developing an implicit PIC
method is the coupling between the equations of motion and the field
equations for the presence of the time advanced electric field (but
not magnetic field, that is used from the previous cycle, as no
instability is introduced) in the equations of motion and for the
 appearance  of the particle properties in the sources
of the Maxwell equations. In both cases the coupling is implicit, so
that the new particle properties need to be known before the fields
can be computed and likewise the new fields need to be available
before the new particle properties can be computed.

\section{Implicit Moment Method}

The implicit moment method  removes the need for iterative methods
and provides a direct method to compute the advanced fields without
first having to move the particles. The implicit moment method reduces the number of equations that
must be solved self-consistently to a set of coupled fluid moment
and field equations.  The solution of these equations implicitly,
and the subsequent solution of the particle equations of
motion in the resulting fields, is stable and accurate.

The coupling due to the implicit discretization of both
field and particle equations is  approximated, representing  the sources
of the field equations using the moment equations instead of the
particle equations directly.  Once the field equations are solved
within this approximation, the rest of the steps can be completed
directly without iterations: with the new fields, the particle
equations of motion can be solved and the new current and density
can be computed for the next computational cycle.

The implicit moment method formulation used here is described in details by  \citet{brackbill-forslund,lapenta06}. The key step is to derive a
suitable set of moment equations that can approximate the particle
motion over a computational cycle. The approach followed in the
present implementation  is based on a
series expansion of the interpolation functions used to transfer information between grid and particles. The details of the simple but demanding algebraic
manipulations are provided by~\citet{vu1992}, the final answer
being:
\begin{equation}
\begin{array}{c}
\rho_s^{n+1}=\rho_s^{n}- \Delta t \nabla \cdot J_s^{n+1/2} \\ \\
\bJ^{n+1/2}_s=\widehat{\bJ}_s-\frac{\Delta t}{2}\boldsymbol{\mu}_s
\cdot E_\theta -\frac{\Delta t}{2} \nabla \cdot
\widehat{\boldsymbol{\Pi}}_s \label{momentimplicit}
\end{array}
\end{equation}
where the following  expressions were defined:
\begin{equation}
\begin{array}{c}
\widehat{\bJ}_s = \sum_p q_p  \widehat{\bv}_p W({\bf x}-{\bf
x}_p^{n})\\ \\ \widehat{\boldsymbol{\Pi}}_s =\sum_p q_p
\widehat{\bv}_p
\widehat{\bv}_p W({\bf x}-{\bf x}_p^{n}) \\ \\
\end{array}
\end{equation}
with the obvious meaning, respectively, of current and pressure
tensor based on the transformed hatted velocities. An effective
dielectric tensor is defined to express the feedback of the electric
field on the plasma current and density:
\begin{equation}
\boldsymbol{\mu}_s^n = - \frac{q_s \rho_s^{n}}{m_s}
\boldsymbol{\alpha}_s^n
\end{equation}

The expression (\ref{momentimplicit}) for the sources of the
Maxwell's equations provide a direct and explicit closure of
Maxwell's equations. When eq.~(\ref{momentimplicit}) is inserted in
eq.~(\ref{differencedMAX}), the Maxwell's equations can be solved
without further coupling with the particle equations. This is the
key property of the moment implicit method and allows the implicit moment 
PIC method to retain the once-through approach typical of explicit
methods and eliminates the need for expensive iteration procedures
that would require to move the particles multiple times per each
computational cycle.

\section{Stability}

The stability properties of the method described above have been
studied extensively in the past~\citep{brackbill-forslund}. All the
stability constraints discussed above for the explicit method are
removed. The implicit particle mover removes the need to resolve the
electron plasma frequency, and the implicit formulation of the field
equations removes the need to resolve the speed of light.

The time step constraints are replaced by an accuracy limit arising
from the derivation of the fluid moment equations using the series
expansion. This limit restricts the mean
particle motion to one grid cell per time
step~\citep{brackbill-forslund}, i.e.
\begin{equation}
\label{time step limit 1} v_{th,e} \Delta t/\Delta x < 1,
\end{equation}
The finite grid instability limit for the explicit method, $\Delta x
< \varsigma \lambda_{De}$ is replaced by~\citep{brackbill-forslund}
\begin{equation}
\Delta x/\Delta t < \varsigma v_{th,e},
\end{equation}
that allows large grid spacings to be used when large time steps are
taken. The gain afforded by the relaxation of the stability limits
is two-fold.

First, the time step can far exceed the explicit limit. In a typical
plasma the electron plasma frequency is far smaller than the time
scales of interest and its accurate resolution is not needed. Within
the current approach, the processes developing at the sub-$\Delta t$
scale are averaged and their energy is damped by a
numerically-enhanced Landau damping. In other approaches, such as
the gyrokinetic or hybrid approach~\citep{multiscale}, such processes
are completely removed and the energy channel towards them is
interrupted, removing for example the possibility to exchange energy
between sub-$\Delta t$ fluctuations and particles. In the implicit
approach, instead, the sub-$\Delta t$ scales remain active and the
energy channel remains open. This is a crucial feature to retain a
full kinetic approach. Furthermore, when additional resolution of
the smallest scales is needed, the implicit method can access the
same accuracy of the explicit method simply using a smaller time
step and grid spacing. This feature is not accessible to reduced
models, e.g. gyroaveraged methods, that remove the small scales
entirely.

Second, the grid spacing can far exceed the Debye length. Often the
scales of interest are much larger than the Debye length. The
ability to retain a full kinetic treatment without the need to
resolve the Debye length results in a much reduced cost for the
implicit PIC method.

\section{Conclusions}

The implicit moment PIC method described above is implemented in the
CELESTE3D code. The CELESTE3D code was originally conceived for the
numerical tokamak project~\citep{itg} but has found its main
application in space physics.

Four types of tests have been conducted to verify and validate
CELESTE in full 3D cartesian geometry and in reduced geometries in
2D and 1D: 1)  well known benchmarks
including shocks, the Weibel instability, Landau damping and ion
acoustic waves~\citep{vu1992}; 2) the GEM challenge~\citep{riccijcp,gem1836} and the Newton challenge~\citep{newton}; 3) study of reconnection in systems with
low betas, investigating the reconnection
process both at the macroscopic and microscopic level obtaining
 agreement with the explicit PIC code NPIC~\citep{gembeta}; 4)  3D stability study of a current sheet equilibrium, compared with satellite observations obtained from the
CLUSTER and GEOTAIL mission~\citep{lapentabrackbill02, lapenta03, riccietal04b}.

\begin{figure}
\centering
 \includegraphics[height=6cm]{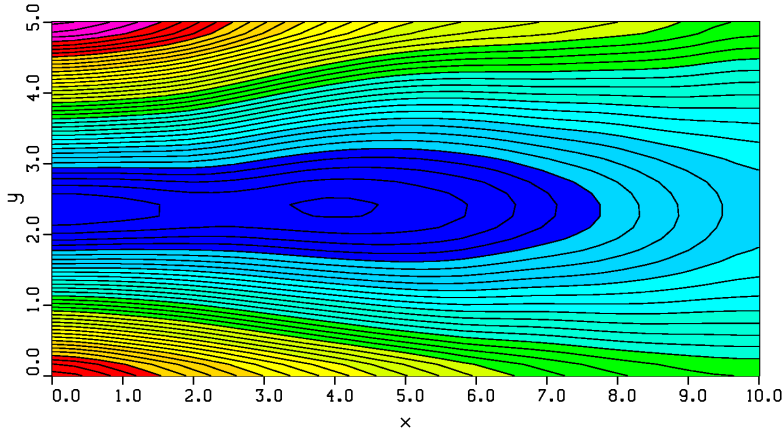}
\caption{3D Simulation of the magnetotail requiring 1 day of CPU time on CELESTE3D or 800,000 years on a explicit code}
\end{figure}

As an example of this last case, we show here the results of an actual simulation conducted in 3D of a system initially in the magnetotail equilibrium described by~\citet{tail-equilibrium}. The physics developing in the simulation includes first the growth and saturation of the lower hybrid drift instability, followed by the onset of reconnection and current flapping leading to the macroscopic restructuring of the topological configuration of the magnetotail. The physics steps of the process are described in deatils by \citet{lapentabrackbill02,lapentanpg,lapenta03, riccietal04b}. 

The fundamental consideration of interest here is the efficacy of the simulation approach. In a typical explicit run, the time step would have to be selected according to the stability constraint of $\omega_{pe}\Delta t<2$. In a typical magnetotail case, the electron plasma frequency is of the order of $\omega_{pe}\approx5\cdot 10^{4}$s$^{-1}$, and the ion plasma frequency is of the order of $\omega_{pi} \approx 10^{3}$s$^{-1}$, both smaller than the smallest scale of interest for this problem which is the lower hybrid frequency range of $\omega_{LH}\approx10^{2}$s$^{-1}$. In a implicit simulation, we can select the time step to the ion plasma frequency, still resolving accurately the lower hybrid range, but saving two orders of magnitude compared with the explicit case that instead is needlessly resolving the electron plasma scale.  

A similar gain occurs in each spatial direction. In a explicit simulation, for stability the grid spacing needs to satisfy $\Delta x/\lambda_{De}<\varsigma$. The Debye length in the magnetotail is of the order of 100m. But the smallest scales of interest in the present case are in the range between the electron (10km) and the ion (100km) inertial scales or gyroscales. In our simulation we set the resolution at 10km, saving two orders of magnitude in each spatial direction but still resolving the important scales. 

Counting all savings, we have saved 2 orders of magnitude in each spatial direction and in time (in total 8 orders of magnitude), without loosing the deatils of the scales of interest. The savings means that on the same computer, an implciti simulation can compute in one day what an explicit simulation would require nearly 800,000 years.

\acknowledgements 
The author is very grateful to Jerry Brackbill and Paolo Ricci for the collaboration in all the work summarized here.
The present work is supported by the {\it Onderzoekfonds K.U. Leuven} (Research Fund KU Leuven), by the European Commission through the SOLAIRE network (MRTN-CT-2006-035484), by the NASA Sun Earth Connection Theory Program and by
 the LDRD program at the
Los Alamos National Laboratory. Work performed in part under the auspices of
the National Nuclear Security Administration of the U.S. Department
of Energy by the Los Alamos National Laboratory, operated by Los
Alamos National Security LLC under contract DE-AC52-06NA25396.
Simulations conducted in part on the
HPC cluster VIC of the Katholieke Universiteit Leuven.



\begin{thebibliography}{19}
\providecommand{\natexlab}[1]{#1}
\providecommand{\url}[1]{{\tt #1}}
\providecommand{\urlprefix}{}
\expandafter\ifx\csname urlstyle\endcsname\relax
  \providecommand{\doi}[1]{doi:\discretionary{}{}{}#1}\else
  \providecommand{\doi}{doi:\discretionary{}{}{}\begingroup
  \urlstyle{rm}\Url}\fi

\bibitem[{Birdsall and Langdon(2004)}]{birdsall}
Birdsall, C. and Langdon, A.: Plasma Physics Via Computer Simulation, Taylor \&
  Francis, London, 2004.

\bibitem[{{Birn}(1987)}]{tail-equilibrium}
{Birn}, J.: {Magnetotail equilibrium theory - The general three-dimensional
  solution}, J. Geophys. Res., 92, 11\,101--11\,108, 1987.

\bibitem[{Birn et~al.(2005)Birn, Galsgaard, Hesse, Hoshino, Huba, Lapenta,
  Pritchett, Schindler, Yin, Buchner, Neukirch, and Priest}]{newton}
Birn, J., Galsgaard, K., Hesse, M., Hoshino, M., Huba, J., Lapenta, G.,
  Pritchett, P.~L., Schindler, K., Yin, L., Buchner, J., Neukirch, T., and
  Priest, E.~R.: Geophys. Res. Lett., 32, L06\,105, 2005.

\bibitem[{Brackbill and Forslund(1982)}]{brackbill-forslund}
Brackbill, J. and Forslund, D.: J. Comp. Phys., 46, 271, 1982.

\bibitem[{Brackbill and Lapenta(1994)}]{itg}
Brackbill, J. and Lapenta, G.: The Effect of Shape on the Ion Temperature
  Gradient Instability, Bull. Am. Phys. Soc., 39, 1665--1666, 1994.

\bibitem[{Denavit(1981)}]{denavit1981}
Denavit, J.: J. Comp. Phys., 42, 337, 1981.

\bibitem[{Hockney and Eastwood(1988)}]{hockney}
Hockney, R. and Eastwood, J.: Computer simulation using particles, Taylor \&
  Francis, London, 1988.

\bibitem[{Langdon et~al.(1983)Langdon, Cohen, and Friedman}]{directimplicit}
Langdon, A., Cohen, B., and Friedman, A.: Direct implicit large time-step
  particle simulation of plasmas, J. Comp. Phys., 51, 107--138, 1983.

\bibitem[{Lapenta and Brackbill(2000)}]{lapentanpg}
Lapenta, G. and Brackbill, J.: Nonlinear Processes Geophys., 7, 151, 2000.

\bibitem[{Lapenta and Brackbill(2002)}]{lapentabrackbill02}
Lapenta, G. and Brackbill, J.: Nonlinear evolution of the lower hybrid drift
  instability: Current sheet thinning and kinking, Phys. Plasmas, 9,
  1544--1554, 2002.

\bibitem[{{Lapenta} et~al.(2003){Lapenta}, {Brackbill}, and
  {Daughton}}]{lapenta03}
{Lapenta}, G., {Brackbill}, J.~U., and {Daughton}, W.~S.: {The unexpected role
  of the lower hybrid drift instability in magnetic reconnection in three
  dimensions}, Phys. Plasmas, 10, 1577--1587, 2003.

\bibitem[{{Lapenta} et~al.(2006){Lapenta}, {Brackbill}, and
  {Ricci}}]{lapenta06}
{Lapenta}, G., {Brackbill}, J.~U., and {Ricci}, P.: {Kinetic approach to
  microscopic-macroscopic coupling in space and laboratory plasmas}, Phys.
  Plasmas, 13, 5904, 2006.

\bibitem[{Lipatov(2002)}]{multiscale}
Lipatov, A. S.~S.: The Hybrid Multiscale Simulation Technology, Springer,
  Berlin, 2002.

\bibitem[{Mason(1981)}]{mason1981}
Mason, R.: J. Comp. Phys., 41, 233, 1981.

\bibitem[{Ricci et~al.(2002{\natexlab{a}})Ricci, Lapenta, and
  Brackbill}]{gem1836}
Ricci, P., Lapenta, G., and Brackbill, J.: Geophys. Res. Lett., 29,
  10.1029/2002GL015\,314, 2002{\natexlab{a}}.

\bibitem[{Ricci et~al.(2002{\natexlab{b}})Ricci, Lapenta, and
  Brackbill}]{riccijcp}
Ricci, P., Lapenta, G., and Brackbill, J.: A Simplified Implicit Maxwell
  Solver, J. Computat. Phys., 183, 117--141, 2002{\natexlab{b}}.

\bibitem[{Ricci et~al.(2004{\natexlab{a}})Ricci, Brackbill, Daughton, and
  Lapenta}]{gembeta}
Ricci, P., Brackbill, J., Daughton, W., and Lapenta, G.: Collisionless magnetic
  reconnection in the presence of a guide field, Phys. Plasmas, 11, 4102--4114,
  2004{\natexlab{a}}.

\bibitem[{Ricci et~al.(2004{\natexlab{b}})Ricci, Lapenta, and
  Brackbill}]{riccietal04b}
Ricci, P., Lapenta, G., and Brackbill, J.: Structure of the magnetotail
  current: Kinetic simulation and comparison with satellite observations,
  Geophys. Res. Lett., 31, L06\,801, doi:10.1029/2003GL019\,207,
  2004{\natexlab{b}}.

\bibitem[{Vu and Brackbill(1992)}]{vu1992}
Vu, H.~X. and Brackbill, J.~U.: Comp. Phys. Comm., 69, 253, 1992.

\end{thebibliography}

\end{document}